\documentclass{Interspeech}

\usepackage{bm}
\usepackage{multirow}
\usepackage{array}
\usepackage{float}
\usepackage[table,xcdraw]{xcolor}
\interspeechcameraready

\title{Exploring Efficient Directional and Distance Cues \\ for Regional Speech Separation}

\author{Yiheng}{Jiang}
\author{Haoxu}{Wang}
\author{Yafeng}{Chen}
\author{Gang}{Qiao}
\author{Biao}{Tian}

\affiliation{Tongyi Lab}{Alibaba Group}{China}
\email{\{jiangyiheng.jyh,tianbiao.tb,chenyafeng.cyf,wanghaoxu.whx,songjiang.qg\}@alibaba-inc.com}

\usepackage{comment}

\begin{document}

\maketitle

\begin{abstract}

In this paper, we introduce a neural network-based method for regional speech separation using a microphone array.
This approach leverages novel spatial cues to extract the sound source not only from specified direction but also within defined distance.
Specifically, our method employs an improved delay-and-sum technique to obtain directional cues, substantially enhancing the signal from the target direction.
We further enhance separation by incorporating the direct-to-reverberant ratio into the input features, enabling the model to better discriminate sources within and beyond a specified distance.
Experimental results demonstrate that our proposed method leads to substantial gains across multiple objective metrics.
Furthermore, our method achieves state-of-the-art performance on the CHiME-8 MMCSG dataset, which was recorded in real-world conversational scenarios, underscoring its effectiveness for speech separation in practical applications.

\end{abstract}

\section{Introduction}

Speech separation, which involves isolating target speech from multiple speakers in noisy environments, is crucial for real-world applications. Advances in this technology greatly improve a range of uses, including selective listening, hearing aids, and conference systems \cite{Nair19-AVZ,Xu22-SVR}.

Current speech separation methods fall into several categories. Some approaches are designed for single-channel separation, starting with deep clustering \cite{hershey2016deep} or permutation invariant training \cite{7952154}.
Other methods utilize personalized characteristics to extract a target speaker's voice and may require an additional module to generate speaker embeddings \cite{Wang19-Voicefilter,Zmolikova19-Speakerbeam}.
In addition, beamforming techniques such as delay-and-sum (DAS) \cite{VanVeen88, VanTrees02-DEM}, minimum variance distortionless response (MVDR) \cite{1326233,erdogan2016mvdr}, and generalized sidelobe canceller (GSC) \cite{Griffiths82,8943308} utilize spatial information to facilitate speech extraction from specific locations.

Recently, deep learning has been increasingly incorporated to improve the robustness of spatial information-based speech separation.
For example, research in \cite{10446587} utilized a neural network to isolate target speech from interference sources across different directional regions using a linear microphone array.
In \cite{Patterson22-DSS}, the authors trained the network to implicitly estimate distance cues to separate speech within a specified distance under noise-free conditions. 
Moreover, a method called Re-Zero \cite{Gu24-ReZero}, captured both directional and distance cues with two cascaded neural networks, respectively, achieving speech separation within a spatial region using a circular microphone array.
Furthermore, there were other studies on direction-based speech separation such as \cite{Xu22-SVR,Wang21-NeuralBeamforming}, while research on distance-based separation remained limited \cite{Patterson22-DSS} due to the challenges of estimating distance cues in noisy and dynamic environments.


\begin{figure}[t]
  \centering
  \includegraphics[width=3.9cm]{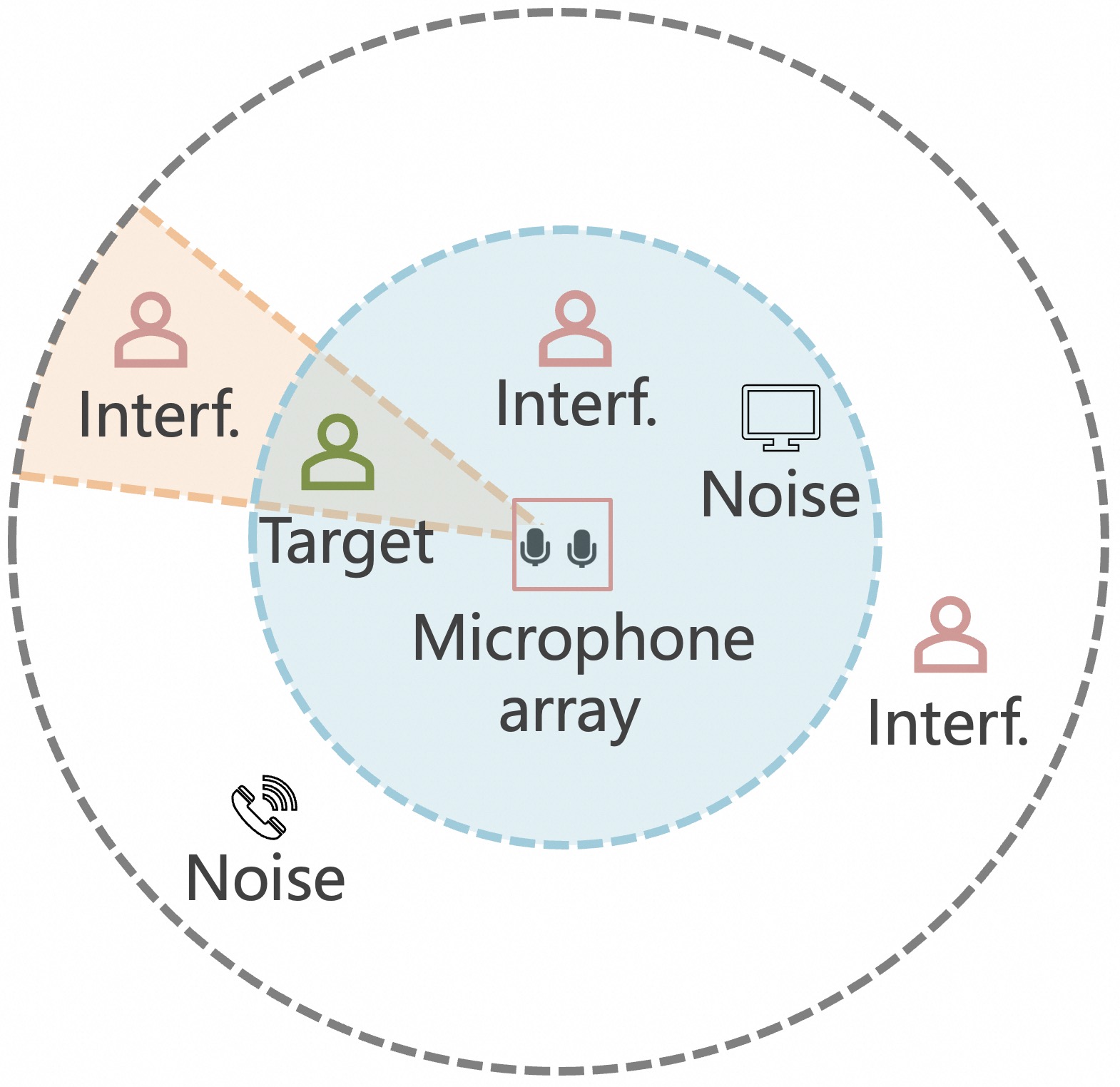}
  \caption{Illustration of regional speech separation. The blue area represents the target distance range, the orange area indicates the target directional range, and their overlap defines the target speech region. ``Interf.'' denotes interference source.}
  \label{fig:fig1}
\end{figure}


Our research, as shown in Figure~\ref{fig:fig1}, focuses on neural network-based regional speech separation (RSS) using a microphone array, where the region is defined by a specified distance and directional range.
While traditional beamforming methods such as DAS achieve only moderate initial signal enhancement in complex environments~\cite{Nugraha2016Multichannel}, they provide crucial directional cues that neural networks can exploit to improve separation performance.
Thus, we utilize an improved DAS approach as an initial filter to optimize directional information in neural network inputs.
Additionally, direct-to-reverberant ratio (DRR) \cite{bronkhorst1999auditory} is inversely proportional to the square of the distance from the talker to the microphones, which provides essential distance information \cite{Blauert1997, Zohourian2020}. By leveraging DRR as a key input feature, we enhance the model's ability to distinguish sound sources within and beyond specified distance. Integrating both DAS and DRR further refines the network's spatial awareness, thereby improving the overall effectiveness of RSS.


Unlike Re-Zero, which requires two independent models for directional and distance discrimination, our solution uses an end-to-end unified model that simultaneously constrains both dimensions for RSS, enhancing computational efficiency and simplifying modeling process.
Moreover, the advantages of our method are validated on the CHiME-8 MMCSG dataset \cite{zmolikova2024chime8},  which features publicly available data recorded in diverse, real-world scenarios using smart glasses, unlike many studies relying on simulated data or controlled conditions \cite{10446587, Patterson22-DSS, Gu24-ReZero, taherian2022location}.

\section{Methods}
\subsection{Problem formulation}

In a known microphone array setup, we tackle the problem of speech separation in scenarios with multiple speakers and background noise. The observed mixture signal is represented as:
\begin{equation}
\bm{y}_i = \sum_{n=1}^{N} \bm{x}_i^n + \bm{v}_i
\label{eq:eq0}
\end{equation}
where \(i \in \{1, 2, \ldots, M\}\) is the index of the microphone.
\(\bm{y}_i \in \mathbb{R}^T\) represents the \(i\)-th channel signal, with \(T\) being the signal length.
$N$ represents the total number of speakers, and \(\bm{v}_i\) captures background noise.
$\bm{x}_i^n$ is the sound received by the $i$-th microphone from the $n$-th speaker, with each speaker potentially located at a different position. 
This study explores novel directional and distance cues for RSS, which aims to extract speech from mixtures within a specified location range.


\subsection{Directional cues}
Although DAS beamforming may not produce entirely satisfactory signals in challenging acoustic conditions, it can serve as a preliminary filtering process that provides clearer directional cues to guide the model learning.
In this work, we build upon the method introduced in \cite{Jenrungrot20-Cone}, which utilizes the ``delay'' step of the DAS beamforming by using a \(shift\) operation to align far-field sources from a specified angle \(\theta\) in the time domain.
\begin{equation}
\bm{y}'_i = shift \left( \bm{y}_i, \left\lfloor (d_{ref}(p_\theta) - d_i(p_\theta)) \cdot \frac{sr}{c} \right\rfloor \right)
\label{eq:eq1}
\end{equation}
where \(d_i(p_\theta)\) and \(d_{ref}(p_\theta)\) denote the distances from position \(p\) at angle \(\theta\) to the \(i\)-th microphone and the reference microphone (typically the first in the setup), respectively.
\(c\) represents sound speed, and \(sr\) is sampling rate.
The floor operation \(\left\lfloor \cdot \right\rfloor\) ensures the computed time delay is an integer, which is then used to shift the microphone signal \(\bm{y}_i\) accordingly.

We extend ``delay'' step by incorporating the ``sum'' operation, completing the DAS beamforming process:
\begin{equation}
  \bm{y}^s = \frac{1}{M}\sum_{i=1}^{M} \bm{y}'_i
  \label{eq:eq2}
\end{equation}
where dividing the summation by \(M\) keeps the output on the same scale as the input signals.
This ``sum'' operation explicitly enhances alignment and emphasizes the target direction.

In addition, we consider the diversity within the ``sum'' step by averaging pairs of signals, providing the model with a richer set of beamforming details to explore spatial cues:
\begin{equation}
\bm{y}_{i,j}^s = \frac{1}{2} (\bm{y}'_i + \bm{y}'_j), \quad \forall i, j \in \{1, 2, \ldots, M\}, \ i \neq j
\label{eq:eq3}
\end{equation}
where \(\bm{y}'_i\) and \(\bm{y}'_j\) represent signals from different microphones, forming a total of \(M(M-1)/2\) pairs.
Moreover, we employ pairwise subtraction to obtain another set of signals (omitting the condition \(\forall i, j \in \{1, 2, \ldots, M\}, \ i \neq j\) for brevity in subsequent discussions):
\begin{equation}
\overline{\bm{y}}_{i,j}^s = \bm{y}'_i - \bm{y}'_j
\label{eq:eq4}
\end{equation}

This subtraction focuses on eliminating aligned target signals, leaving outputs with more pronounced interference and other components.
This allows the model to better understand the patterns of non-target parts.


The signal set $\bm{y}_{das} = \{\bm{y}'_i, \bm{y}^s, \bm{y}_{i,j}^s, \overline{\bm{y}}_{i,j}^s\}$ is transformed into 80-dimensional Fbank features, referred to as $\bm{Y}_{das}$, which consists of $M^2 + 1$ channels and serves as DAS features for directional discrimination.
Increasing the number of microphones expands the signal pairs, significantly raising the channel count of \(\bm{Y}_{das}\) and potentially overwhelming processing capacity.
However, our experiments show that using just a subset of signal pairs is sufficient to achieve effective separation.

\subsection{Distance cues}

Theoretically, DRR encompasses rich distance information, allowing for the prediction of distance in both single-channel \cite{bronkhorst1999auditory} and multi-channel \cite{vesa2009binaural} scenarios.
Intuitively, DRR also functions as a reliable cue for training neural networks to perform distance-based speech separation.
Inspired by \cite{Lu2010}, we employ short-time Fourier transform (STFT) features of the aligned signal pairs \((\bm{y}'_i, \bm{y}'_j)\), denoted as \((\bm{Y}_i(t, f), \bm{Y}_j(t, f))\), to compute the DRR, where \((t, f)\) represents time and frequency indices. For simplicity, \((t, f)\) will be omitted in subsequent discussions.

First, we calculate the gain of \(\bm{Y}_i\) with respect to \(\bm{Y}_j\):
\begin{equation}
  \bm{G}_{i,j} = \frac{|\bm{Y}_j|}{|\bm{Y}_i| + \epsilon}
\end{equation}
where \(\epsilon\) is a small positive value for numerical stability.
The gain \(\bm{G}_{i,j}\) then serves as compensation factor to adjust the amplitude of \(\bm{Y}_i\) using element-wise multiplication:
\begin{equation}
  \bm{Y}_i^g = \bm{Y}_i \odot \bm{G}_{i,j}
\end{equation}
Since the time delay of alignment in Equation (\ref{eq:eq1}) is calculated based on the source's direct path, we follow \cite{Lu2010} in assuming that the direct sound in \(\bm{Y}_j\) matches that in the compensated signal \(\bm{Y}_i^g\). Consequently, the residual energy is calculated as:
\begin{equation}
  \bm{R}_{i,j} = \left| \bm{Y}_j - \bm{Y}_i^g \right|^2
\end{equation}
where $\bm{R}_{i,j}$ includes energy from reverberation, misaligned interference sources and noise.
The aligned direct sound energy is then determined by:
\begin{equation}
  \bm{D}_{i,j} = \left|\bm{Y}_j\right|^2 - \bm{R}_{i,j}
\end{equation}

As reverberant energy is part of \(\bm{R}_{i,j}\), we incorporate both direct sound energy \(\bm{D}_{i,j}\) and residual energy \(\bm{R}_{i,j}\) into the DRR features \(\bm{Y}_{drr} = \{\bm{D}_{i,j}, \bm{R}_{i,j}\}\) for distance discrimination. We explore using either the concatenation \([\bm{D}_{i,j}, \bm{R}_{i,j}]\) or the decibel (dB) ratio \(10\log_{10}(\bm{D}_{i,j}/\bm{R}_{i,j})\) to assist the model learning. In concatenation mode, the model is expected to discern the relationship between direct sound and reverberant components. In ratio mode, \(\bm{R}_{i,j}\) acts as a proxy for reverberation, directly providing the DRR form to the model for distance separation.

\section{Experimental setup}
\subsection{System implementation}
Multiple directional ranges can be regions of interest in our implementation for greater flexibility. However, to ensure alignment with a specified direction, only one source is designated as the target in each training sample. This direction, represented by \(\theta\) in Equation \eqref{eq:eq1}, is determined by the central angle of the directional region to which the target belongs. During inference, the sample can be realigned to various angles, allowing predictions for different directional regions.



The RSS model inputs consist of DAS features \(\bm{Y}_{das}\) and DRR features \(\bm{Y}_{drr}\), employed as directional and distance cues, respectively.
To address the differing scales of \(\bm{Y}_{das}\) and \(\bm{Y}_{drr}\), batch normalization (BN) is applied to each of them individually. Then, fully connected (FC) layers are used to align their dimensions.
The backbone of the model is a dual-path recurrent neural network (DPRNN) \cite{9054266} composed of four blocks. Each block integrates two long short-term memory (LSTM) layers \cite{hochreiter1997long} that operate over temporal and frequency dimensions, respectively, facilitating streaming inference.
The dimension of the hidden state in LSTM is 48.
The model consists of 979K parameters, and for a 2-second speech input, the FLOPs is 6.2G.
The model outputs a mask \(\bm{m}\) applied to the STFT of \(\bm{y}^s\), thereby predicting the signal through \(\text{iSTFT}(\bm{m} \odot \text{STFT}(\bm{y}^s))\), where \(\text{iSTFT}\) denotes the inverse STFT.
The training criterion is Modulation Loss~\cite{Vuong2021Modulation}.
The sampling rate of audio signals is 16 KHz.
The entire inference process incurs only a 20 ms latency, primarily due to feature transformation using 40 ms frame lengths and 20 ms frame shifts.

\subsection{Data preparation}
We first conducted experiments on simulated data. Clean speech and point noise clips came from the DNS challenge~\cite{Reddy2020}, and isotropic noise was sourced from \cite{7953152}.
Room impulse responses (RIRs) were generated for an 8-channel linear microphone array with a diameter of 38 cm using gpuRIR \cite{diaz2020gpuRIR}.
The microphone array was randomly positioned within rooms whose dimensions varied from \(3 \times 3 \times 2.5\) meters to \(10 \times 8 \times 4\) meters.
We defined two target speech regions within 1.8 meters, each with distinct azimuth ranges \([70^\circ\),\(80^\circ]\) and \([100^\circ\),\(110^\circ]\).
We omitted elevation angles, as this simplification aligns with many real-world applications \cite{Gu24-ReZero,Jenrungrot20-Cone}.
For each simulated sample, in addition to placing a source within a specified target region, a point noise source was randomly positioned within the room. Moreover, we generated three interfering speech sources in relation to the target source, as depicted in Figure \ref{fig:fig1}: 

\begin{itemize} [left=2em] 
\item \(Interf(a)\): Same azimuth range, outside distance, 
\item \(Interf(b)\): Within distance, outside azimuth range, 
\item \(Interf(c)\): Outside both azimuth range and distance. 
\end{itemize}

Besides the simulated test dataset, we also validated our method on the CHiME-8 MMCSG dataset. It contains multimodal data, but we focused exclusively on audio signals. It comprises two-party conversations recorded in real-world environments using smart glasses with an irregular array of 7 microphones.
The recordings capture sounds from the wearer, talking partner, interfering talker and background noise.
This dataset also provides additional RIRs, which we used for training RSS model.
More details are available in \cite{zmolikova2024chime8}.

In our experiments, both the wearer and the partner are treated as target sources, with one of them randomly selected as the target during RSS model training.
The regions of these two targets were determined based on RIRs statistics.
Note that in addition to azimuth and distance, the elevation angle is also crucial in this glasses scenario, as it provides clear differentiation between the two target identities due to the head-mounted setup.
Through parallel streaming inference, which aligns with two target regions separately, the total latency of the RSS algorithm is 20 ms.
Combined with a streaming automatic speech recognition (ASR) algorithm running concurrently on both target audios from RSS outputs, our testing was conducted in a fully streaming real-time fashion.

\section{Results}
\begin{table}[]
  \center
  \caption{Performance comparison of DAS features on direction-based speech separation.} \label{tab0}
  \tabcolsep=0.127cm
  \renewcommand{\arraystretch}{1.1}
  \begin{tabular} {l|cccc} 
  \hline
  Inputs                      & PESQ & STOI(\%) & SDR(dB) & Decay(dB) \\ \hline
                              Raw-Mics                         & 1.67                         & 79.1                         & 6.18                        & 26.9                          \\
                              Aligned-Mics                   & 1.81                         & 83.7                         & 7.12                        & 32.0                          \\
                              DAS                     & 2.47                         & 91.5                         & 10.06                       & 49.7                          \\
  \textbf{DAS(all pairs)} & \textbf{2.50}                & \textbf{92.0}                & \textbf{10.18}              & \textbf{55.6}                 \\ \hline
  \end{tabular}
\end{table}
\begin{table}[]
  \center
  \caption{Performance comparison of DRR features on distance-based speech separation.} \label{tab1}
  \tabcolsep=0.16cm
  \renewcommand{\arraystretch}{1.1}
  \begin{tabular}{l|cccc} 
  \hline
  Inputs                      & PESQ & STOI(\%) & SDR(dB) & Decay(dB) \\ \hline
                              Raw-Mics                         & 1.83                         & 82.4                         & 7.38                        & 35.6                          \\
                              Aligned-Mics                   & 1.83                         & 82.9                         & 7.54                        & 39.3                          \\
                              DRR(cat)                & 1.91                         & 84.0                         & 7.98                        & 42.8                          \\
                              \textbf{DRR(ratio)}             & \textbf{2.19}                & \textbf{88.7}                & \textbf{9.12}               & \textbf{49.2}                 \\ \hline
  \end{tabular}
\end{table}
\subsection{Directional separation}

We first performed direction-based speech separation without considering distance, allowing us to assess the directional impact before adding distance in future experiments.
In this setup, both the target source and \(Interf(a)\) are considered as targets because they are from the same azimuth range, whereas \(Interf(b)\) and \(Interf(c)\) are classified as interference sources. By randomly selecting some of these sources and noise to synthesize a mixed signal, we created a test dataset that includes several cases: \((a)\) For samples without a target source, we use Decay~\cite{Gu24-ReZero} to measure energy reduction in the separated output relative to the mixture. \((b)\) For single-target samples, we evaluate performance using perceptual evaluation of speech quality (PESQ) \cite{Rix2001} and short-term objective intelligibility (STOI) \cite{Taal2010}. \((c)\) For samples with one or two targets, we apply signal-to-distortion ratio (SDR) \cite{LeRoux2019} for assessment.

In Table \ref{tab0}, \emph{Raw-Mics} involves using raw 8-microphone signals for separation.
\emph{Aligned-Mics} employs the ``delay'' step to align multi-channel signals before feeding them to the network. Building on this, \emph{DAS(all pairs)} further applies our proposed DAS features, utilizing all possible signal pairs and resulting in 65 input channels.
In contrast, the basic \emph{DAS} experiment uses only 17 channels from four symmetric positional pairs (0,7), (1,6), (2,5), and (3,4) within the linear microphone array.

The \emph{Aligned-Mics} experiment efficiently enhances directional separation. However, our \emph{DAS} implementation achieves significant improvements that far surpass the \emph{Aligned-Mics} experiment across all metrics, clearly demonstrating the superior effectiveness of the proposed DAS features. Additionally, the \emph{DAS(all pairs)} approach offers only minor benefits compared to the \emph{DAS} experiment, suggesting that a subset of signal pairs is sufficient for robust separation. This setup prevents explosive growth in DAS feature channels due to excessive microphones.

\subsection{Distance separation}
We now focus on experiments involving distance separation. In this setup, the target source and \(Interf(b)\) are potential targets as they are both within the specified distance.
As illustrated in Table \ref{tab1}, we incorporate distance distinction using the proposed DRR features in two modes:
\emph{DRR(cat)} for concatenation mode and \emph{DRR(ratio)} for ratio mode.
Compared to the \emph{Aligned-Mics} experiment, the \emph{DRR(cat)} provides moderate improvements, while \emph{DRR(ratio)} demonstrates substantial enhancements, highlighting the advantage of ratio mode over concatenation mode.
Another observation from comparing Tables \ref{tab0} and \ref{tab1} is that distance separation is somewhat less effective than directional separation in metrics, due to directional cues offering stronger discrimination than distance cues \cite{Gu24-ReZero}.

\begin{table}[]
  \center
  \caption{Performance comparison of regional features on RSS.} \label{tab2}
  \tabcolsep=0.075cm
  \renewcommand{\arraystretch}{1.1}
  \begin{tabular}{l|cccc} 
  \hline
  Inputs                      & PESQ & STOI(\%) & SDR(dB) & Decay(dB) \\ \hline
                              Mixture                      & 1.12                         & 61.1                         & -3.15                       & 0.0                           \\
                              Direction model          & 1.50                         & 73.9                         & 3.34                        & 24.1                          \\
                              Distance model      & 1.47                         & 72.2                         & 3.45                        & 27.2                          \\ \hline
                              Raw-Mics                         & 1.64                         & 76.2                         & 5.75                        & 46.6                          \\
                              DAS                     & 1.93                         & 85.0                         & 7.85                        & 71.1                          \\
                              DRR(ratio)                 & 1.90                         & 83.5                         & 7.43                        & 54.4                          \\
                              \textbf{DAS+DRR(ratio)}             & \textbf{2.16}                & \textbf{88.2}                & \textbf{8.79}               & \textbf{75.7}                 \\ \hline
  \end{tabular}
\end{table}
\begin{table}[]
  \center
  \caption{Impact of reverberation on region-based speech separation, based on DAS+DRR(ratio) experiment.} \label{tab10}
  \tabcolsep=0.15cm
  \renewcommand{\arraystretch}{1.1}
  \begin{tabular}{l|cccc}
  \hline
  T60 range      & PESQ & STOI(\%) & SDR(dB) & Decay(dB) \\ \hline
  0.05s$\ \sim\ $0.3s & 2.11 & 88.19 & 8.67 & 77.08 \\
  0.3s$\ \sim\ $0.55s & 2.15 & 88.05 & 8.80 & 74.26 \\
  0.55s$\ \sim\ $0.8s & 2.20 & 88.49 & 8.95 & 75.76 \\
  0.8s$\ \sim\ $1.1s  & 2.14 & 88.30 & 8.79 & 76.50 \\
  1.1s$\ \sim\ $1.4s  & 2.12 & 87.40 & 8.53 & 74.95 \\ \hline
  \end{tabular}
\end{table}
\begin{figure}[t]
  \centering
  \includegraphics[width=5cm]{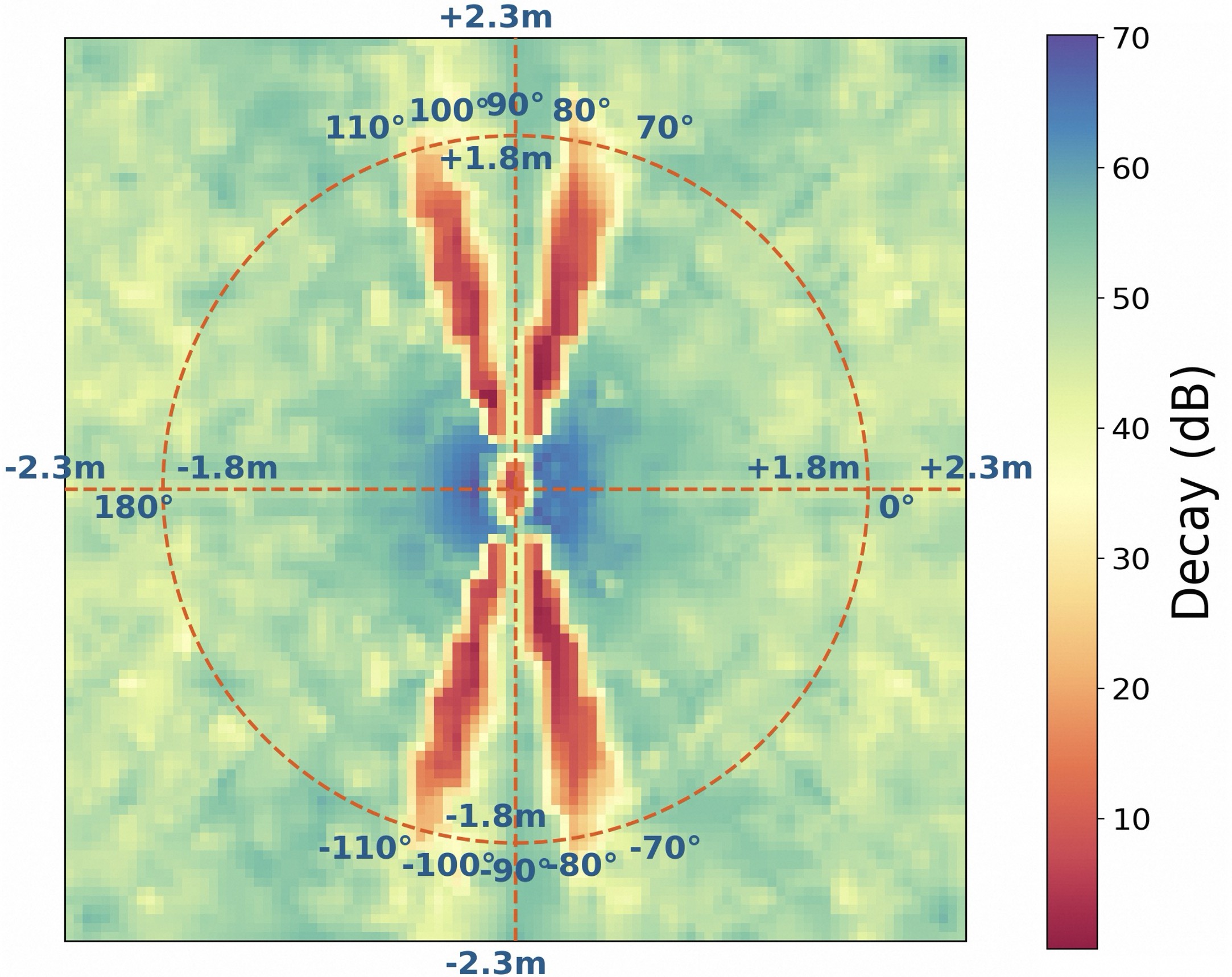}
  \caption{Heatmap of Decay at various coordinates in a room based on DAS+DRR(ratio) experiment.}
  \label{fig:fig2}
\end{figure}

\subsection{Regional separation}
Next, we analyze the RSS experiment results, which incorporate both directional and distance distinctions.
Unlike previous studies, the RSS experiments involve only one target source, with \(Interf(a)\), \(Interf(b)\) and \(Interf(c)\) all viewed as interference.
As shown in Table \ref{tab2}, the baseline \emph{Mixture} uses the mixture signal $\bm{y}^s$ as the separation output.
The \emph{Direction model} is from the \emph{DAS} experiment in Table \ref{tab0}, and the \emph{Distance model} is from the \emph{DRR(ratio)} experiment in Table \ref{tab1}.


While the \emph{Direction model} and \emph{Distance model} show some improvement over the baseline \emph{Mixture}, their performance is constrained as they primarily suppress specific interference sources, leaving others inadequately addressed.
Shifting our focus to the last four rows, the \emph{DAS} and \emph{DRR(ratio)} experiments independently introduce directional and distance cues for RSS training, yielding notable gains over the \emph{Raw-Mics} experiment.
Combining these cues in \emph{DAS+DRR(ratio)} experiment further enhances separation results, achieving an STOI of 88.2 and an SDR of 8.79, thereby confirming the reliability of our approach for RSS.
Additionally, Decay reaches a large value of 75.7, possibly due to the suppression of numerous interfering components (up to three interferers).
This increased interference presents additional challenges in our simulated test dataset, setting it apart from other works such as \cite{Gu24-ReZero}.


Since reverberation can degrade speech quality \cite{wang2020multi} and affect DRR calculations, we assessed its impact on our method under varying reverberation strengths using the \emph{DAS+DRR(ratio)} experiment.
As shown in Table \ref{tab10}, the metrics change only slightly with different reverberation times (T60). The T60 in the training dataset ranges from 0.05 to 0.8 s, and even when it exceeds 0.8 s in the test dataset, the performance metrics remain stable. This demonstrates the capability of our method to manage challenging reverberation conditions.



For better visualization, we simulated a \(5 \times 5 \times 3\) meters room with the microphone array centrally located. A sound source is placed at various coordinates on the horizontal \(xy\)-plane at \(z = 1.5\) meters, keeping at least 20 cm away from the walls, resulting in a \(4.6 \times 4.6\) meters area. Based on the model of the \emph{DAS+DRR(ratio)} experiment, we tested the Decay of the separation outputs as shown in \figurename~\ref{fig:fig2}.
Our method accurately distinguishes sound sources located at azimuth ranges of \([70^\circ, 80^\circ]\) and \([100^\circ, 110^\circ]\) within a distance of 1.8 meters. Due to front-back confusion in linear arrays \cite{Jenrungrot20-Cone}, the figure also indicates corresponding separable angles on the opposite side.
\begin{table}[]
  \center
  \caption{WER (\%) results on CHiME-8 MMCSG eval dataset.} \label{tab3}
  \tabcolsep=0.03cm
  \renewcommand{\arraystretch}{1.1}
  \begin{tabular}{ll|cccc}
  \hline
  \multicolumn{1}{l}{Latency range}                            & \ System        & All            & Wearer         & Partner        & Latency(s)              \\ \hline
  \multirow{3}{*}{0s$\ \sim\ $0.15s}                & baseline~\cite{zmolikova2024chime8}      & 22.10          & 17.80          & 26.30          & 0.138          \\
                                                    & NPU-TEA~\cite{huang2024npu_tea}       & 16.80          & 12.20          & 21.30          & 0.142          \\
                                                    & \textbf{Ours} & \textbf{15.20} & \textbf{10.50} & \textbf{19.90} & \textbf{0.127} \\ \hline
  \multirow{3}{*}{0.15s$\ \sim\ $0.35s} & baseline~\cite{zmolikova2024chime8}      & 18.90          & 15.00          & 22.90          & \textbf{0.333}          \\
                                                    & NPU-TEA~\cite{huang2024npu_tea}       & 15.00          & 10.80          & 19.20          & 0.337          \\
                                                    & \textbf{Ours} & \textbf{13.65} & \textbf{9.30}  & \textbf{18.00} & 0.343 \\ \hline
  \multirow{3}{*}{0.35s$\ \sim\ $1s}    & baseline~\cite{zmolikova2024chime8}      & 17.90          & 14.10          & 21.70          & 0.616          \\
                                                    & FOSAFER~\cite{huang2024fossafer}       & 14.00          & 9.60           & 18.40          & 0.686          \\
                                                    & \textbf{Ours} & \textbf{13.05} & \textbf{8.90}  & \textbf{17.20} & \textbf{0.587} \\ \hline
  \end{tabular}
\end{table}
\subsection{MMCSG testing}

To evaluate the effectiveness of our method in real-world scenarios, we conducted streaming ASR experiments on the CHiME-8 MMCSG glasses dataset under three latency range conditions, as shown in Table \ref{tab3}.
The evaluation metric is word error rate (WER) for two speakers: the \emph{Wearer} and the \emph{Partner}, with \emph{All} representing their average.
As our solution conducts parallel inferences for both identities, the total latency includes a 20 ms delay from the RSS algorithm, with the remainder attributed to the ASR model.

The results demonstrate that our method consistently outperforms state-of-the-art systems such as NPU-TEA and FOSAFER across all latency range conditions.
Notably, we followed the baseline approach (see~\cite{zmolikova2024chime8}), using only about 8 hours of data collected with glasses to train the ASR model, whereas other systems leveraged extensive open-source datasets for ASR optimization.
Additionally, unlike other solutions requiring the ASR model to recognize both identities simultaneously using serialized-output-training~\cite{Kanda2022}, our method processes each speaker independently.
This strategy enhances the coherence and naturalness of the recognition context, resulting in a notable improvement compared to other systems.

\section{Conclusion}
We have developed a novel neural network-based approach that incorporates multi-faceted spatial cues for RSS.
By integrating an improved DAS beamforming for directional enhancement and utilizing a reliable DRR technique for distance distinction, our method achieves significant improvements across multiple metrics.
The effectiveness demonstrated on the CHiME-8 MMCSG dataset highlights our method's potential for real-world conversational applications, particularly in scenarios requiring precise selective listening with low computational latency.

\bibliographystyle{IEEEtran}
\bibliography{mybib}

\begin{thebibliography}{10}
\providecommand{\url}[1]{#1}
\csname url@samestyle\endcsname
\providecommand{\newblock}{\relax}
\providecommand{\bibinfo}[2]{#2}
\providecommand{\BIBentrySTDinterwordspacing}{\spaceskip=0pt\relax}
\providecommand{\BIBentryALTinterwordstretchfactor}{4}
\providecommand{\BIBentryALTinterwordspacing}{\spaceskip=\fontdimen2\font plus
\BIBentryALTinterwordstretchfactor\fontdimen3\font minus \fontdimen4\font\relax}
\providecommand{\BIBforeignlanguage}[2]{{%
\expandafter\ifx\csname l@#1\endcsname\relax
\typeout{** WARNING: IEEEtran.bst: No hyphenation pattern has been}%
\typeout{** loaded for the language `#1'. Using the pattern for}%
\typeout{** the default language instead.}%
\else
\language=\csname l@#1\endcsname
\fi
#2}}
\providecommand{\BIBdecl}{\relax}
\BIBdecl

\bibitem{Nair19-AVZ}
A.~A. Nair, A.~Reiter, C.~Zheng, and S.~Nayar, ``Audiovisual zooming: What you see is what you hear,'' in \emph{Proc. ACM International Conference on Multimedia (ACMMM)}, 2019, pp. 1107--1118.

\bibitem{Xu22-SVR}
A.~Xu and R.~R. Choudhury, ``Learning to separate voices by spatial regions,'' in \emph{Proc. International Conference on Machine Learning (ICML)}, 2022, pp. 24\,539--24\,549.

\bibitem{hershey2016deep}
J.~R. Hershey, Z.~Chen, J.~Le~Roux, and S.~Watanabe, ``Deep clustering: Discriminative embeddings for segmentation and separation,'' in \emph{Proc. ICASSP}, 2016, pp. 31--35.

\bibitem{7952154}
D.~Yu, M.~Kolbæk, Z.-H. Tan, and J.~Jensen, ``Permutation invariant training of deep models for speaker-independent multi-talker speech separation,'' in \emph{Proc. ICASSP}, 2017, pp. 241--245.

\bibitem{Wang19-Voicefilter}
Q.~Wang, H.~Muckenhirn, K.~Wilson, P.~Sridhar, Z.~Wu, J.~R. Hershey, R.~A. Saurous, R.~J. Weiss, Y.~Jia, and I.~L. Moreno, ``Voicefilter: Targeted voice separation by speaker-conditioned spectrogram masking,'' in \emph{Proc. Interspeech}, 2019, pp. 2728--2732.

\bibitem{Zmolikova19-Speakerbeam}
K.~Zmolikova, M.~Delcroix, K.~Kinoshita, T.~Ochiai, T.~Nakatani, L.~Burget, and J.~Cernocky, ``Speakerbeam: Speaker aware neural network for target speaker extraction in speech mixtures,'' \emph{IEEE Journal of Selected Topics in Signal Processing}, vol.~13, no.~4, pp. 800--814, 2019.

\bibitem{VanVeen88}
B.~D.~V. Veen and K.~M. Buckley, ``Beamforming: A versatile approach to spatial filtering,'' \emph{IEEE ASSP Magazine}, vol.~5, no.~2, pp. 4--24, April 1988.

\bibitem{VanTrees02-DEM}
H.~L.~V. Trees, \emph{Detection, Estimation, and Modulation Theory, Part IV: Optimum Array Processing}.\hskip 1em plus 0.5em minus 0.4em\relax New York: Wiley, 2002.

\bibitem{1326233}
C.~Richmond, ``The {CAPON-MVDR} algorithm: Threshold {SNR} prediction and the probability of resolution,'' in \emph{Proc. ICASSP}, vol.~2, 2004, pp. 217--220.

\bibitem{erdogan2016mvdr}
H.~Erdogan, J.~R. Hershey, S.~Watanabe, M.~I. Mandel, and J.~L. Roux, ``Improved {MVDR} beamforming using single-channel mask prediction networks,'' in \emph{Interspeech 2016}, 2016.

\bibitem{Griffiths82}
L.~J. Griffiths and C.~W. Jim, ``An alternative approach to linearly constrained adaptive beamforming,'' \emph{IEEE Transactions on Antennas and Propagation}, vol.~30, no.~1, pp. 27--34, January 1982.

\bibitem{8943308}
D.~Chang and B.~Zheng, ``Adaptive generalized sidelobe canceler beamforming with time-varying direction-of-arrival estimation for arrayed sensors,'' \emph{IEEE Sensors Journal}, vol.~20, no.~8, pp. 4403--4412, 2020.

\bibitem{10446587}
Y.~Yang, G.~Sung, S.~Shih, H.~Erdogan, C.~Lee, and M.~Grundmann, ``Binaural angular separation network,'' in \emph{Proc. ICASSP}, 2024, pp. 1201--1205.

\bibitem{Patterson22-DSS}
K.~Patterson, K.~Wilson, S.~Wisdom, and J.~R. Hershey, ``Distance-based sound separation,'' in \emph{Proc. INTERSPEECH}, 2022, pp. 901--905.

\bibitem{Gu24-ReZero}
R.~Gu and Y.~Luo, ``{ReZero}: Region-customizable sound extraction,'' \emph{IEEE/ACM Transactions on Audio, Speech, and Language Processing}, vol.~32, pp. 2576--2589, 2024.

\bibitem{Wang21-NeuralBeamforming}
Z.~Wang, H.~Erdogan, S.~Wisdom, K.~Wilson, D.~Raj, S.~Watanabe, Z.~Chen, and J.~R. Hershey, ``Sequential multi-frame neural beamforming for speech separation and enhancement,'' in \emph{IEEE Spoken Language Technology Workshop (SLT)}, 2021.

\bibitem{Nugraha2016Multichannel}
A.~A. Nugraha, A.~Liutkus, and E.~Vincent, ``Multichannel audio source separation with deep neural networks,'' \emph{IEEE/ACM Transactions on Audio, Speech, and Language Processing}, vol.~24, no.~9, pp. 1652--1662, September 2016.

\bibitem{bronkhorst1999auditory}
A.~W. Bronkhorst and T.~Houtgast, ``Auditory distance perception in rooms,'' \emph{Nature}, vol. 397, no. 6719, pp. 517--520, February 1999.

\bibitem{Blauert1997}
J.~Blauert, \emph{Spatial Hearing: The Psychophysics of Human Sound Localization}.\hskip 1em plus 0.5em minus 0.4em\relax MIT Press, 1997.

\bibitem{Zohourian2020}
M.~Zohourian and R.~Martin, ``Binaural direct-to-reverberant energy ratio and speaker distance estimation,'' \emph{IEEE/ACM Transactions on Audio, Speech, and Language Processing}, vol.~28, pp. 92--104, 2020.

\bibitem{zmolikova2024chime8}
K.~Zmolikova, S.~Merello, K.~Kalgaonkar, J.~Lin, N.~Moritz, P.~Ma, M.~Sun, H.~Chen, A.~Saliou, S.~Petridis, C.~Fuegen, and M.~Mandel, ``The {CHiME-8 MMCSG} challenge: Multi-modal conversations in smart glasses,'' in \emph{8th International Workshop on Speech Processing in Everyday Environments (CHiME 2024)}, September 2024, pp. 7--11.

\bibitem{taherian2022location}
H.~Taherian, K.~Tan, and D.~Wang, ``Location-based training for multi-channel talker-independent speaker separation,'' in \emph{Proc. ICASSP}.\hskip 1em plus 0.5em minus 0.4em\relax IEEE, 2022.

\bibitem{Jenrungrot20-Cone}
T.~Jenrungrot, V.~Jayaram, S.~M. Seitz, and I.~Kemelmacher-Shlizerman, ``The cone of silence: Speech separation by localization,'' in \emph{Proc. Conference on Neural Information Processing Systems (NeurIPS)}, 2020.

\bibitem{vesa2009binaural}
S.~Vesa, ``Binaural sound source distance learning in rooms,'' \emph{IEEE Transactions on Audio, Speech, and Language Processing}, vol.~17, no.~8, pp. 1498--1507, November 2009.

\bibitem{Lu2010}
Y.~C. Lu and M.~Cooke, ``Binaural estimation of sound source distance via the direct-to-reverberant energy ratio for static and moving sources,'' \emph{IEEE Transactions on Audio, Speech, and Language Processing}, vol.~18, no.~7, pp. 1793--1804, September 2010.

\bibitem{9054266}
Y.~Luo, Z.~Chen, and T.~Yoshioka, ``Dual-path {RNN}: Efficient long sequence modeling for time-domain single-channel speech separation,'' in \emph{Proc. ICASSP}, 2020, pp. 46--50.

\bibitem{hochreiter1997long}
S.~Hochreiter and J.~Schmidhuber, ``Long short-term memory,'' \emph{Neural computation}, vol.~9, no.~8, pp. 1735--1780, 1997.

\bibitem{Vuong2021Modulation}
T.~Vuong, Y.~Xia, and R.~Stern, ``A modulation-domain loss for neural-network-based real-time speech enhancement,'' in \emph{Proc. ICASSP}, 2021.

\bibitem{Reddy2020}
C.~K. Reddy, V.~Gopal, R.~Cutler, E.~Beyrami, and R.~Cheng, ``The interspeech 2020 deep noise suppression challenge: datasets, subjective testing framework, and challenge results,'' in \emph{Proc. Interspeech}, 2020, pp. 340--354.

\bibitem{7953152}
T.~Ko, V.~Peddinti, D.~Povey, M.~L. Seltzer, and S.~Khudanpur, ``A study on data augmentation of reverberant speech for robust speech recognition,'' in \emph{Proc. ICASSP}, 2017, pp. 5220--5224.

\bibitem{diaz2020gpuRIR}
D.~Diaz-Guerra, A.~Miguel, and J.~Beltran, ``{gpuRIR}: A python library for room impulse response simulation with gpu acceleration,'' \emph{Multimedia Tools and Applications}, vol.~79, no.~24, pp. 17\,275--17\,297, 2020.

\bibitem{Rix2001}
A.~W. Rix, J.~G. Beerends, M.~P. Hollier, and A.~P. Hekstra, ``Perceptual evaluation of speech quality ({PESQ})-a new method for speech quality assessment of telephone networks and codecs,'' in \emph{Proc. ICASSP}, vol.~2.\hskip 1em plus 0.5em minus 0.4em\relax IEEE, 2001, pp. 749--752.

\bibitem{Taal2010}
C.~H. Taal, R.~C. Hendriks, R.~Heusdens, and J.~Jensen, ``A short-time objective intelligibility measure for time-frequency weighted noisy speech,'' in \emph{Proc. ICASSP}.\hskip 1em plus 0.5em minus 0.4em\relax IEEE, 2010, pp. 4214--4217.

\bibitem{LeRoux2019}
J.~L. Roux, S.~Wisdom, H.~Erdogan, and J.~R. Hershey, ``{SDR}--half--baked or well done?'' in \emph{Proc. ICASSP}.\hskip 1em plus 0.5em minus 0.4em\relax IEEE, 2019, pp. 626--630.

\bibitem{wang2020multi}
Z.~Q. Wang and D.~L. Wang, ``Multi-microphone complex spectral mapping for speech dereverberation,'' in \emph{Proc. ICASSP}, 2020, pp. 486--490.

\bibitem{huang2024npu_tea}
K.~Huang, W.~Rao, Y.~Li, H.~Wang, Y.~Wang, S.~Huang, and L.~Xie, ``The {NPU-TEA} system report for the {CHiME-8} {MMCSG} challenge,'' in \emph{CHiME Workshop on Speech Processing in Everyday Environments}, 2024.

\bibitem{huang2024fossafer}
S.~Huang \emph{et~al.}, ``The {FOSAFER} system for the {CHiME-8} {MMCSG} challenge,'' in \emph{CHiME Workshop on Speech Processing in Everyday Environments}, 2024.

\bibitem{Kanda2022}
N.~Kanda, J.~Wu, Y.~Wu, X.~Xiao, Z.~Meng, X.~Wang, Y.~Gaur, Z.~Chen, J.~Li, and T.~Yoshioka, ``Streaming speaker-attributed asr with token-level speaker embeddings,'' in \emph{Proceedings of Interspeech 2022}, 2022, pp. 521--525.

\end{thebibliography}

\end{document}